\documentclass[a4paper]{jpconf}
\pdfoutput=1
\usepackage{iopams}
\usepackage{graphicx}

\begin{document}
\title[Impact of packing fraction on pattern formation]{Impact of packing fraction on diffusion-driven pattern formation in a two-dimensional system of rod-like particles}

\author{Yuri~Yu~Tarasevich$^1$, Valeri~V~Laptev$^{1,2}$, Valentiva V Chirkova$^1$, Nikolai~I~Lebovka$^{3,4}$}
\address{$^1$Astrakhan State University, 20A Tatishchev Street, Astrakhan, 414056, Russia}
\address{$^2$Astrakhan State Technical University, 16 Tatishchev Street, Astrakhan, 414025, Russia}
\address{$^3$F. D. Ovcharenko Institute of Biocolloidal Chemistry, NAS of Ukraine, 42 Boulevard Vernadskogo, 03142 Kiev, Ukraine}
\address{$^4$Taras Shevchenko Kiev National University, Department of Physics, 64/13 Volodymyrska Street, 01601 Kiev, Ukraine}
\ead{tarasevich@asu.edu.ru}

\begin{abstract}
Pattern formation in a two-dimensional system of rod-like particles has been simulated using a lattice approach. Rod-like particles were modelled as linear $k$-mers of two mutually perpendicular orientations ($k_x$- and $k_y$-mers) on a square lattice with periodic boundary conditions (torus). Two kinds of random sequential adsorption model were used to produce the initial homogeneous and isotropic distribution of $k$-mers with different values of packing fraction.
By means of the Monte Carlo technique, translational diffusion of the $k$-mers was simulated as a random walk, while rotational diffusion was ignored, so, $k_x$- and $k_y$-mers were considered as individual species. The system tends toward a well-organized nonequilibrium steady state in the form of diagonal stripes for the relatively long $k$-mer ($k \geq 6$) and moderate packing densities (in the interval $p_{down} < p < p_{up}$, where both the critical packing fractions $p_{down}$  and $p_{up}$ are depended on $k$).
\end{abstract}

\section{Introduction}\label{sec:intro}
The various complex nonequilibrium phase behavior, orientational ordering, and self-organization have been experimentally observed in vibrated systems composed of shape-anisotropic (particularly, rod-like) particles in two dimensions~\cite{Narayan2006,Galanis2006,Aranson2007,Galanis2010,Yadav2012EPJE,Boerzsoenyi2013,Muller2015PRE,Gonzalez2017SM}. The experiments demonstrated that, at high packing fraction, the shape of the particles is important for the orientational ordering. Numerous examples of patterns and of phase behavior in granular media together with appropriate references can be found in the review~\cite{Aranson2006RMP}.

In recent decades, much attention has been paid to the study of systems of linear $k$-mers (particles occupying $k$ adjacent adsorption sites) deposited on 2D lattices. A linear $k$-mer represents the simplest model of an hard-core (completely rigid) rod-like particle with an aspect ratio of $k$. Computer simulations have been extensively applied to investigate percolation and jamming phenomena for the random sequential adsorption (RSA)~\cite{Evans1993RMP} of $k$-mers (see, e.g.,~\cite{Centres2015JStatMech,Kuriata2016,Budinski2017PRE} and the references therein).

Recently, diffusion-driven pattern formation in a 2D system of $k$-mers has been studied by means of Monte Carlo (MC) simulation~\cite{Lebovka2017PRE,Tarasevich2017JSM}. The $k$-mers were oriented in two mutually perpendicular directions. For those systems with equal number of species belonging to two different kinds, i.e., $k_x$-mers and $k_y$-mers,
\begin{enumerate}
  \item nonequilibrium steady patterns in the form of stripe domains have been observed only for periodic boundary conditions along both directions (PBCs);
  \item this self-organization was possible only in fairly dense systems;
  \item these stripe domains have been observed only for relatively long $k$-mer, $k \geqslant 6$;
  \item relaxation time to the well-organized  steady-state depends drastically on lattice size.
\end{enumerate}
Besides, self-organization can occur in the systems with unequal numbers of $k_x$-mers and $k_y$-mers.

Nevertheless,  effect of packing fraction (i.e., surface coverage) on such pattern formation has not been studied in detail.
Particularly, jammed states have been produced in~\cite{Lebovka2017PRE,Tarasevich2017JSM} using RSA.
Notice, that other mechanisms, e.g., RSA with diffusional relaxation can produce denser systems~\cite{Fusco2001JChPh}. It is not clear whether self-organization possible in such the dense systems?

Present conference paper is devoted to detailed analysis how packing fraction affects the pattern formation in a two-species diffusion system. For this reason, we examine the effect of packing fraction on the pattern formation using two kinds of RSA to produce initial homogeneous and isotropic state.

The rest of the paper is organized as follows. In section~\ref{sec:methods}, the technical details of the simulations are described and all necessary quantities are introduced. Section~\ref{sec:results} presents our principal findings. Section~\ref{sec:conclusion} summarizes the main results.

\section{Details of simulation}\label{sec:methods}
In our study, a lattice approach is used and the problem was simulated using a square lattice of size $L\times L$. All calculations were performed using only one lattice size $L=256$. This choice  is due to the fact that relaxation time increases drastically as lattice size increases~\cite{Lebovka2017PRE,Tarasevich2017JSM}. We used toroidal boundary conditions, i.e., periodic boundary conditions along both the $x$ and $y$ axes (PBCs).

The rod-like particles were presented as linear $k$-mers. The previous data evidenced that self-organization is not observed for short $k$-mers ($k<6$)~\cite{Lebovka2017PRE}] and in this work the length of the $k$-mers was varied from 6 to 12. Longer particles were excluded from consideration due to expected essential finite-size effect (ratio $L/k$ is too small for $k>12$). Isotropic deposition of the  $k$-mers was ensured, i.e., $k$-mers oriented along the $x$ and $y$ directions ($k_x$-mers and $k_y$-mers, respectively) were equiprobable in their deposition. This corresponded to the zero value of a mean order parameter of the system, defined as
\begin{equation}\label{eq:s}
s = \frac{\left|N_y - N_x\right|}{N},
\end{equation}
where $N_x$ and $N_y$  are the numbers of sites occupied by $k_x$-mers and $k_y$-mers, respectively, and $N = N_y + N_x$ is the total number of occupied sites.

The diffusion of $k$-mers was simulated using the kinetic MC procedure. In our simulation, only translational diffusion was taken into consideration. This is essentially the case for fairly dense systems in the jamming state, where rotational diffusion is impeded, especially for large values of~$k$. Undoubtedly, for dilute systems, rotational diffusion can occur, however, this was ignored in our study.
An arbitrary $k$-mer was randomly chosen at each step and a translational shift by one lattice unit along either the longitudinal or the transverse axis of the $k$-mer was attempted.
All the four possible directions to shift the $k$-mer were attempted in a random order until a direction is found in which the displacement of the particle is possible, or until all possible directions have been exhausted. Note, that this kinetics can violate detailed balance condition and drives the system to a nonequilibrium steady state~\cite{Patra2017}. Nevertheless, this kinetics is natural for ``intellectual'' particles, e.g., biological species which are looking for life resources or pedestrians.
One time step of the MC computation, which corresponds to an attempted displacement of the total number of $k$-mers in the system, $N$, was taken as the MC time unit. Time  counting was started from the value of $t_{MC}=0$, being the initial moment (before diffusion), and the total duration of the simulation was $10^7$ MC time units.

The RSA model~\cite{Evans1993RMP} was used to produce an initial homogeneous distribution of linear $k$-mers. These $k$-mers were deposited randomly and sequentially, and their overlapping with previously placed particles was forbidden, i.e. excluded volume interaction was assumed. The packing fraction, $p$, varies in the range  $p \in [0.1, p_j]$, where $p_j$ is the packing fraction at jamming. In jammed state, no additional $k$-mer can be placed because the presented voids are too small or of inappropriate shape. To avoid confusion, here and below, we will use $p_j$ for packing fraction of the jammed state produced only by RSA.

\begin{figure}[htb]
\begin{minipage}[c]{0.65\textwidth}
 \centering
  \includegraphics[width=\textwidth]{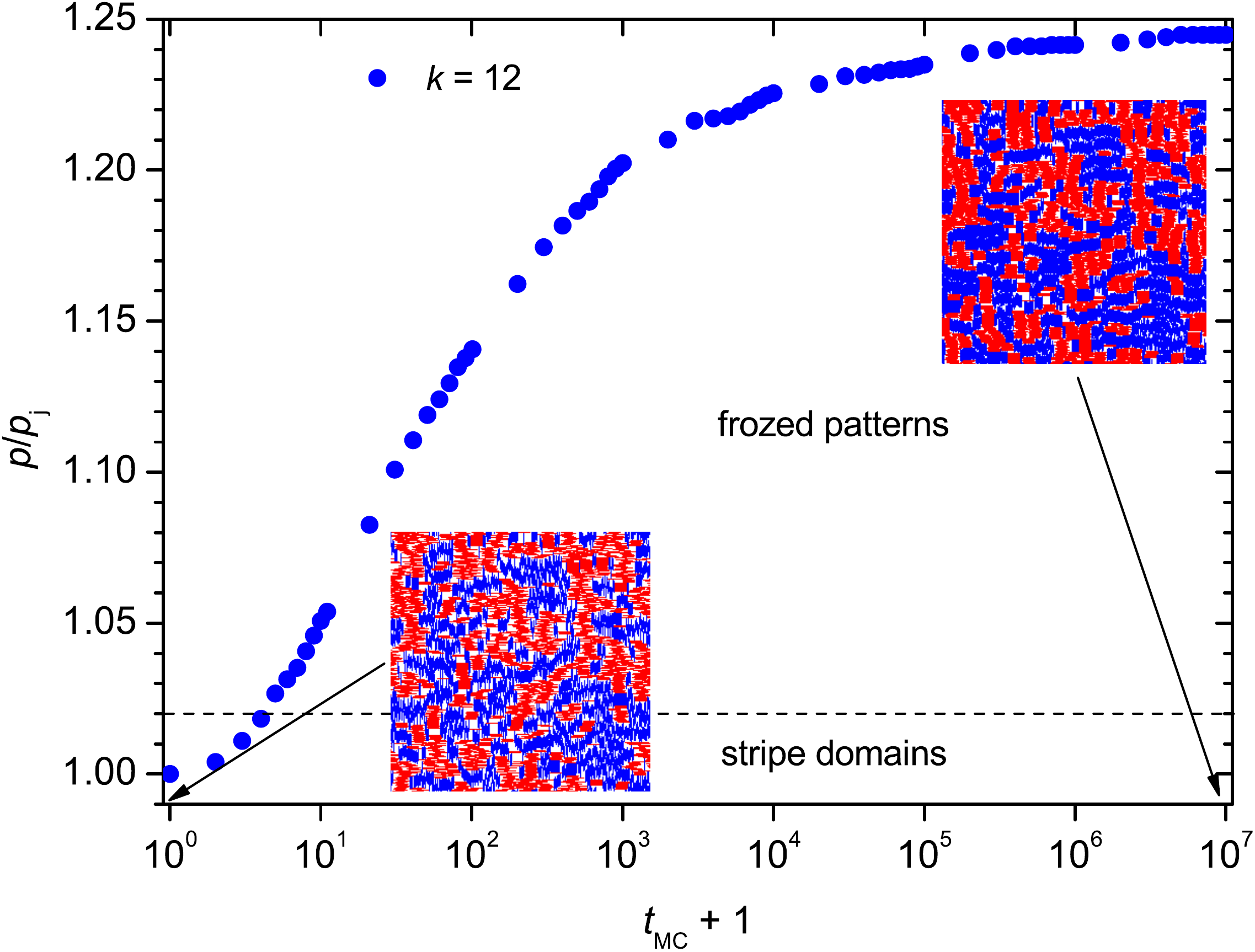}
\end{minipage}\hfill
\begin{minipage}[c]{0.3\textwidth}
  \caption{Growth of packing fraction with time, $k=12$. Inserts demonstrate initial jammed state produced using RSA ($p_j = 0.72$, $t_{MC} =0$) and final high-density state obtained using extended RSA approach ($p=0.89 \approx1.25p_j$, $t_{MC} = 10^7$). Notice, that some more $k$-mers still can be added to the latter state even after $10^7$ MC steps.}\label{fig:densityvstime}
\end{minipage}
\end{figure}

We additionally studied the effect of packing fraction on self-organization using the extended RSA approach for obtaining the packing fractions $p>p_j$. To produce such the dense state, we used two-step algorithm.
\begin{description}
 \item[At the first step,] RSA was used to produce a jammed state.
 \item[At the second step,] the particles were allowed to diffuse. Due to diffusion, some additional voids can occur. When such the void is large enough, one additional particle was obligatory deposited onto the lattice. This action resembles shaking a sugar bowl when we want to add some more sugar in it. 
     This additional deposition was stopped when the packing fraction reached the necessary value.
\end{description}
     Figure~\ref{fig:densityvstime} demonstrates how packing fraction increases with time for one particular case. 
     In the case of unlimited additional deposition, fairly dense packing fraction $p \approx 1.25p_j$ can be reached for $6 \leq k \leq 12$. Additional particle deposition leads to suppression of self-organization. The preliminary studies shown that the initial structure is actually frozen, but becomes more and more dense during  additional deposition of the particles.

After reaching the desirable packing fraction, only diffusion of particles is allowed.
We will use abbreviation RSADA for this process of the extended RSA with diffusion and additional adsorption.

To characterize the temporal evolution of the system under consideration, several quantities were monitored at each given MC step.
\begin{enumerate}
\item  The normalized number of clusters, $n$, i.e., the current value divided by the value at the initial state, $t_{MC} = 0$. The clusters built of $k_x$-mers and clusters built of $k_y$-mers have been accounted separately and then averaged. We used the Hoshen--Kopelman algorithm~\cite{HK76} to count the clusters.
    \item The local anisotropy, $s$, i.e., the order parameter $s$ calculated using \eref{eq:s} in a window of $l \times l$ sites and averaged over the entirely set of windows. We used $l = 2^{-n} L$, $n = 1,2,\dots,5$. The principal analysis has been performed for $l = L/4$. For this window size, we calculated ratios of final value of $s$ at $t_{MC}=10^7$ to its initial value at $t_{MC}=0$.
   \item The fraction of interspecific contacts $n_{xy}^{\ast} = n_{xy} / ( n_{xy} + n_x + n_y )$, where $n_{xy}$ is the number of interspecific contacts between the different sorts of $k$-mers (i.e., $k_x$--$k_y$), $n_x$ and $n_y$  are the numbers of intraspecific contacts between $k$-mers of the same kind (i.e., $k_x$--$k_x$ and $k_y$--$k_y$, respectively). For convenience of comparison, we additionally used ratios of final value of $n_{xy}^{\ast}$ at $t_{MC}=10^7$ to its initial value at $t_{MC}=0$.
  \item The shift ratio, $R$, i.e., the ratio of the number of shifts of the $k$-mers along the transverse axes to the number of shifts along their longitudinal axes during one MC step.
  \item The electrical conductivities, $\sigma$. We used the method described in~\cite{Lebovka2017PRE} to transform the system under consideration into a random resistor network (RRN). The Frank--Lobb algorithm~\cite{Frank1988PRB} was applied to calculate the electrical conductivity of such RRNs, (see figure~\ref{fig:conduct} for the details of calculations).
\end{enumerate}
All these quantities were averaged over 100 independent statistical runs for each pair $(p,k)$.

\section{Results}\label{sec:results}
\begin{figure}[!hbp]
  \centering\small
($a$)\,\includegraphics[width=0.15\linewidth]{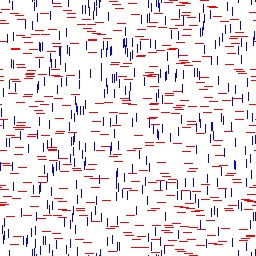}\, ($b$)\,\includegraphics[width=0.15\linewidth]{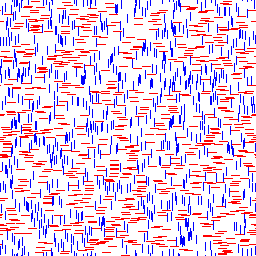}\,
($c$)\,\includegraphics[width=0.15\linewidth]{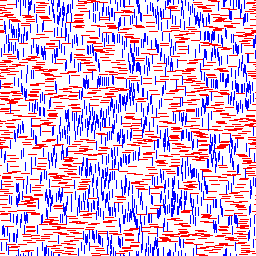}\,
($d$)\,\includegraphics[width=0.15\linewidth]{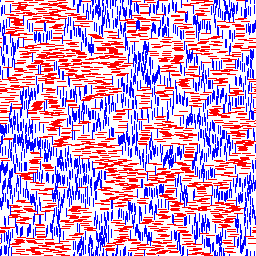}\,
($e$)\,\includegraphics[width=0.15\linewidth]{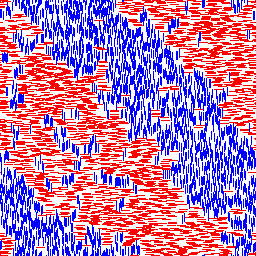}\\
($f$)\,\includegraphics[width=0.15\linewidth]{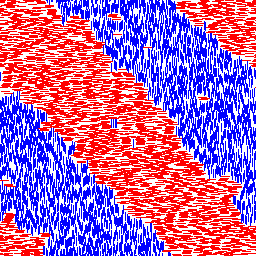}\,
($g$)\,\includegraphics[width=0.15\linewidth]{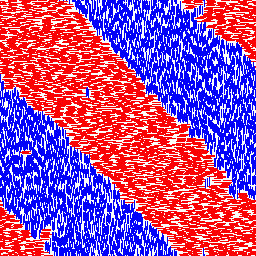}\,
($h$)\,\includegraphics[width=0.15\linewidth]{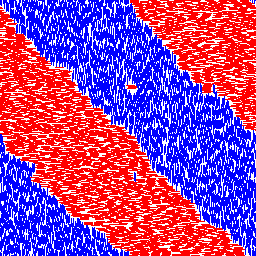}\,
($i$)\,\includegraphics[width=0.15\linewidth]{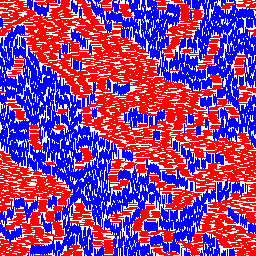}\,
($j$)\,\includegraphics[width=0.15\linewidth]{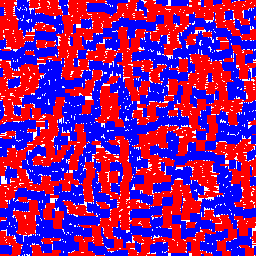}\\
\caption{Examples of systems at $t_{MC}=10^7$ for different values of packing fractions, $p$. $k=9$, $L=256$. ($a$)~$p=0.1$, ($b$)~$p=0.2$, ($c$)~$p=0.3$, ($d$)~$p=0.4$, ($e$)~$p=0.5$, ($f$)~$p=0.6$, ($g$)~$p=0.7$, ($h$)~$p=p_j \approx 0.743$, ($i$)~$p=0.75$, ($j$)~$p=0.925$.}\label{fig:patternsconcentrations}
\end{figure}

We examined the effect of packing fraction, $p$, on pattern formation. Figure~\ref{fig:patternsconcentrations} presents examples of final patterns ($t_{MC} = 10^7$)  at different values of $p$ for $k=9$. Up to $p=0.4$, any regular patterns do not occur despite visible rearrangement of the system (figure~\ref{fig:patternsconcentrations}($a$)--($d$)). Then, stripe domains form when the value of packing fraction, $p$, varies from $p_{down} = 0.5$  up to jamming concentration, $p_j$, (figure~\ref{fig:patternsconcentrations}($e$)--($h$)). Slightly above $p_j$ ($p \approx 1.02p_j$), stripe domains do not form during $10^7$ MC steps (figure~\ref{fig:patternsconcentrations}($i$)), nevertheless, rearrangement of the system is still continuing. For larger values of $p$, mobility of $k$-mers is essentially restricted, namely, a $k$-mer can perform small longitudinal irregular oscillations with zero mean square displacement (figure~\ref{fig:patternsconcentrations}($j$)).

\begin{figure}[!htbp]
  \centering
  \includegraphics[width=\textwidth]{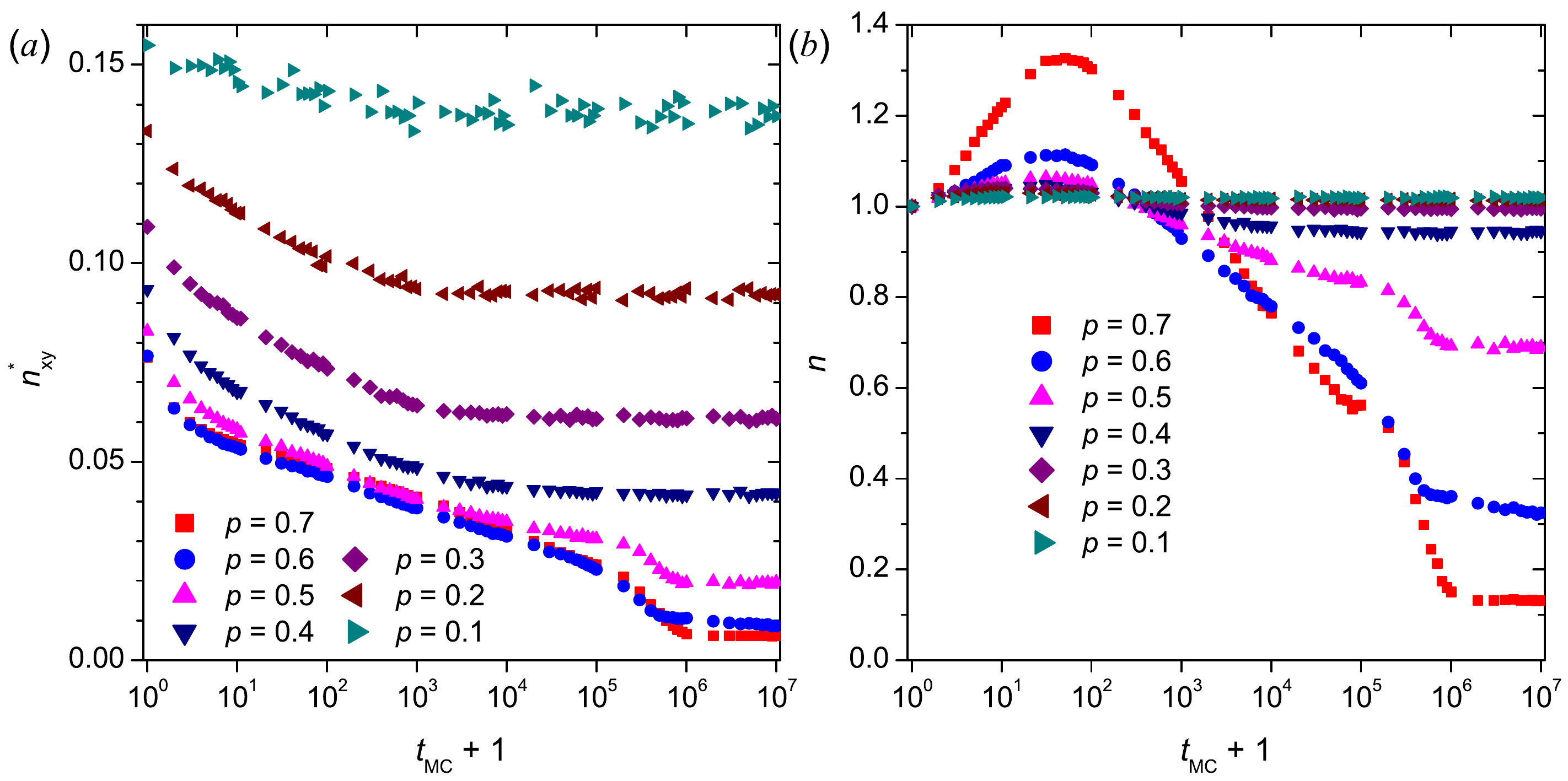}
  \caption{Examples of dependencies of relative number of contacts, $n^*_{xy}$, ($a$) and normalized number of clusters, $n$, ($b$) vs MC steps, $t_{MC}$, for different values of packing fraction, $p$, and fixed value of $k$, $k=9$.}\label{fig:nnxyk9}
\end{figure}

Figure~\ref{fig:nnxyk9} demonstrates some examples of variations of relative number of contacts, $n^*_{xy}$, ($a$) and normalized number of clusters, $n$, ($b$) vs MC steps, $t_{MC}$, for different values of packing fraction, $p$, and fixed value of $k$, $k=9$. Packing fractions $p=0.5, 0.6, 0.7$ correspond to pattern formation whereas no pattern formation was observed for smaller values of packing fractions. Both $n^*_{xy}\left(t_{MC}\right)$ and $n\left(t_{MC}\right)$ curves have a stepwise decrease between $t_{MC}= 10^5$ and $t_{MC}= 10^6$ when packing fractions correspond to pattern formation.

\begin{figure}[!htbp]
\begin{minipage}[c]{0.6\textwidth}
  \centering
  \includegraphics[width=\textwidth]{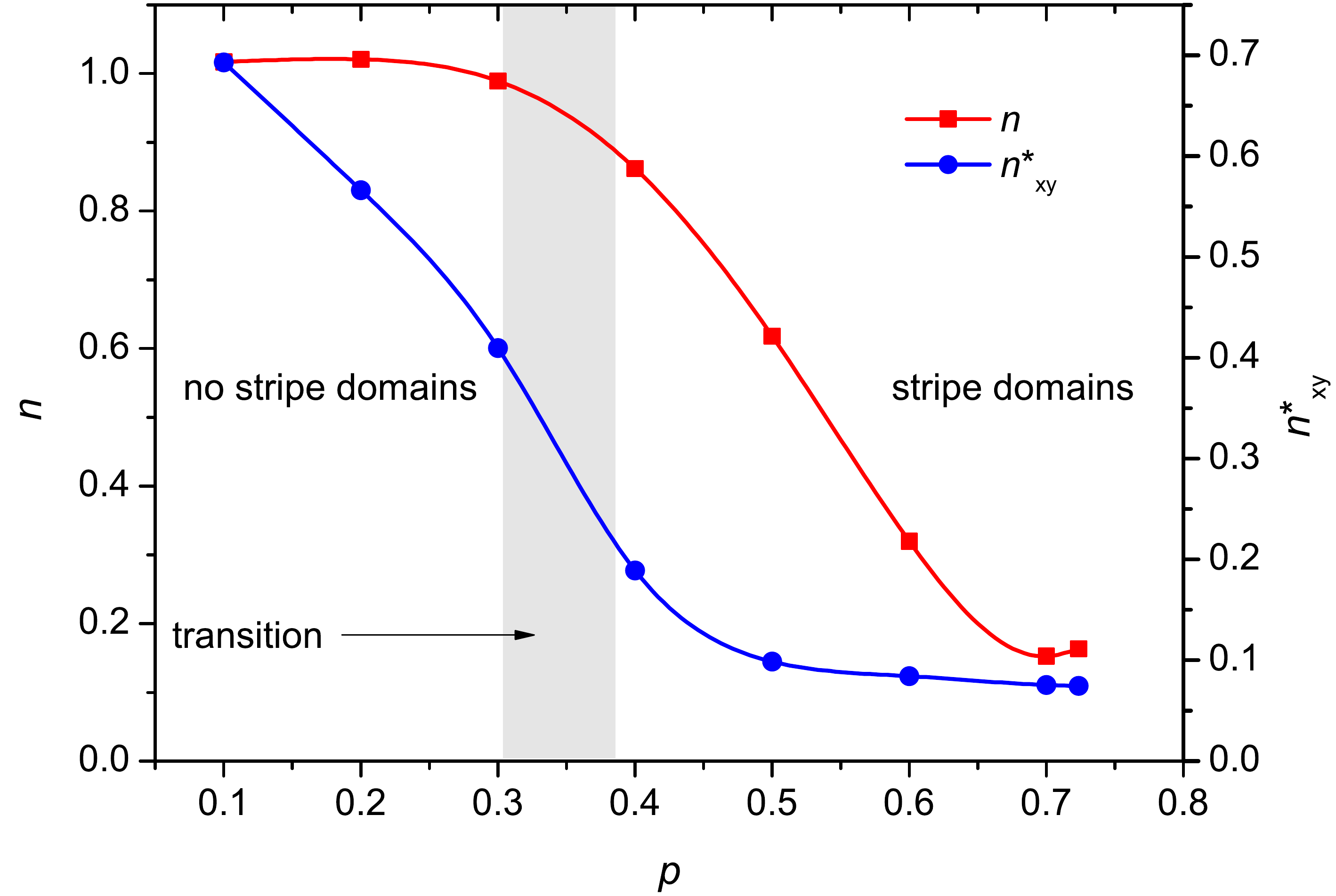}
  \end{minipage}\hfill
\begin{minipage}[c]{0.35\textwidth}
  \caption{Examples of dependencies of relative number of contacts, $n^*_{xy}$, and normalized number of clusters, $n$, vs the packing fraction, $p$, for $k=12$.}\label{fig:nnxyvsp}
  \end{minipage}
\end{figure}

Figure~\ref{fig:nnxyvsp} demonstrates how relative number of contacts, $n^*_{xy}$, and normalized number of clusters, $n$, varies with the packing fraction, $p$, for $k=12$. The occurrence of stripe domains is not accompanied by any characteristic change in the curves. Hence, these quantities are hardly helpful for monitoring of self-organization and its characterization.

\begin{figure}[!htbp]
\begin{minipage}[c]{0.6\textwidth}
  \centering
  \includegraphics[width=\textwidth]{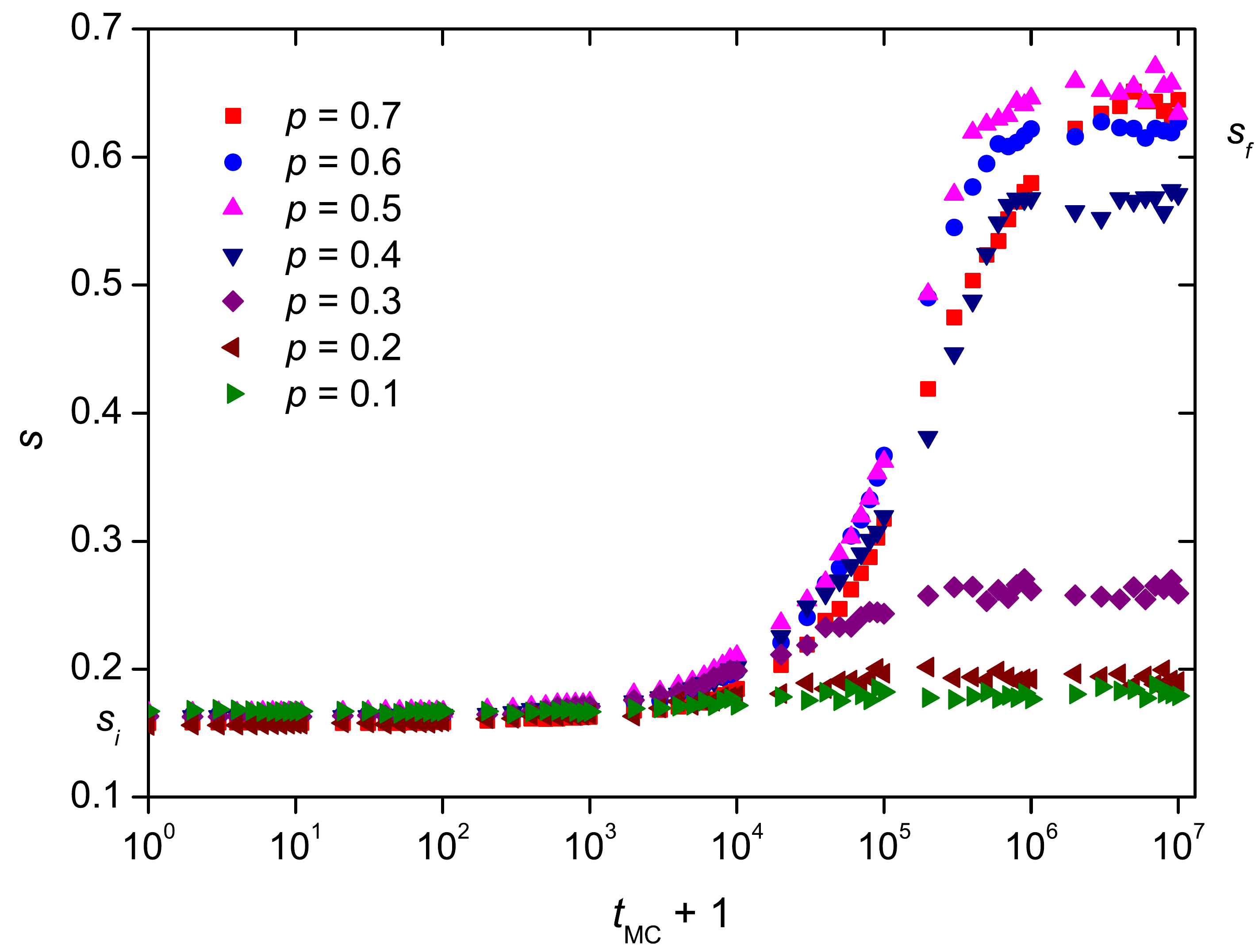}
    \end{minipage}\hfill
\begin{minipage}[c]{0.35\textwidth}
  \caption{Examples of dependencies of local order parameter, $s$, calculated in windows of size $64 \times 64$ sites vs MC steps, $t_{MC}$, for different values of packing fraction, $p$, and fixed value of $k$, $k=12$. $s_i = s(0)$, $s_f = s\left(10^7\right)$.}\label{fig:sl64k12}
    \end{minipage}
\end{figure}

In contrast, local order parameter offers the clear evidence of long-range pattern formation (figure~\ref{fig:sl64k12}). Essential growth of the order parameter after $10^5$ MC steps is a result of cluster coarsening. This effect is pronounced for packing fractions $p=0.5, 0.6, 0.7$ and almost inappreciable for smaller values of packing fractions.

\begin{figure}[!htbp]
\begin{minipage}[c]{0.6\textwidth}
  \centering
  \includegraphics[width=\textwidth]{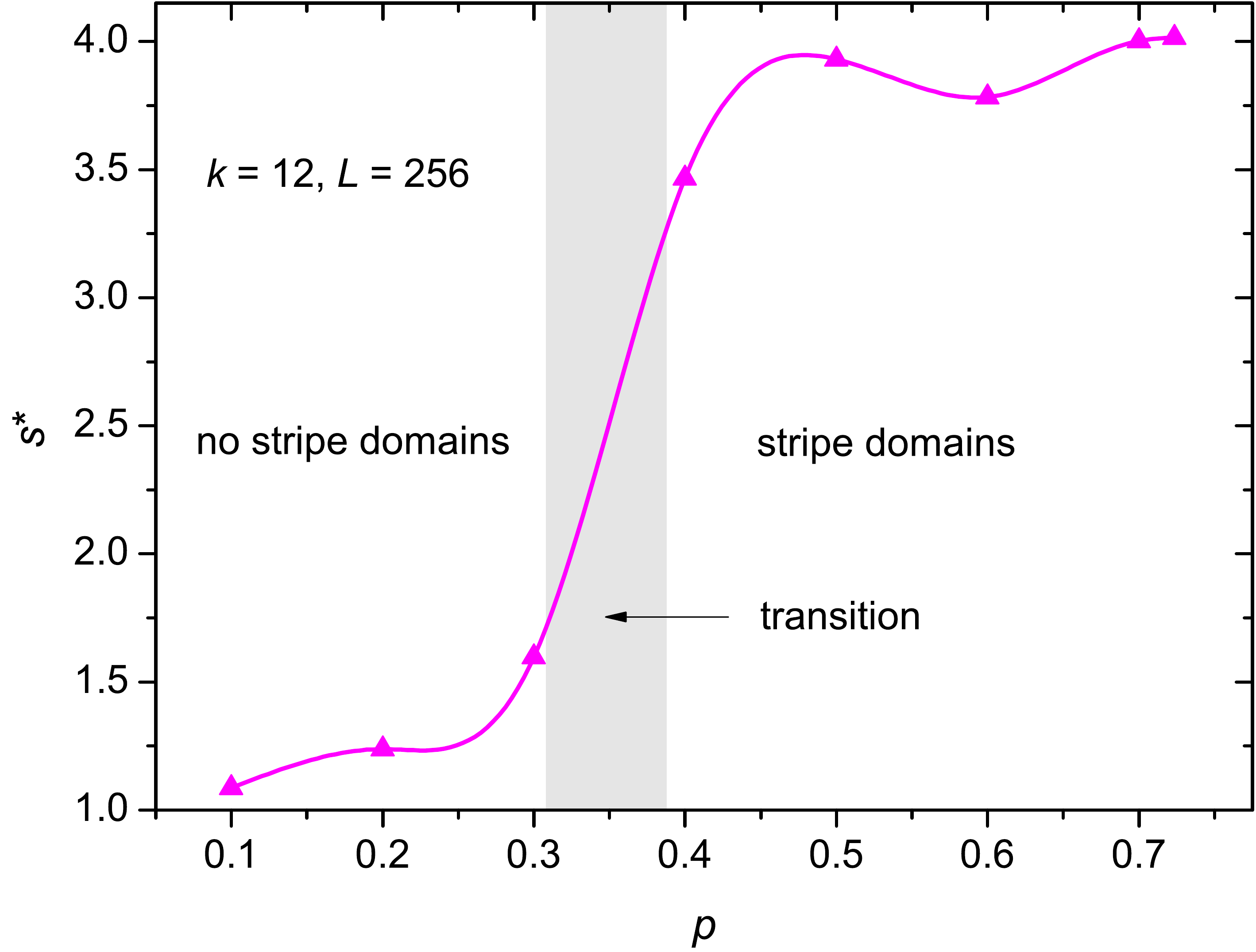}
      \end{minipage}\hfill
\begin{minipage}[c]{0.35\textwidth}
  \caption{Example of dependency of relative local order parameter, $s^*= s_f/s_i$, vs packing fraction, $p$, for $k=12$.}\label{fig:svsp}
  \end{minipage}
\end{figure}

A transition from a homogeneous steady-state to a patterned one looks like continuous, i.e., the stripe domains become less pronounced when the initial packing fraction decreases. At some initial packing fractions, the steady-state looks quite homogeneous (figure~\ref{fig:patternsconcentrations}($d$)) whereas imperfect stripe domains start to form at larger initial packing fractions (figure~\ref{fig:patternsconcentrations}($e$)). All the quantities of interest change smoothly as the packing fraction changes (figure~\ref{fig:nnxyk9}). However, near the critical packing fraction, $p_{down}$, changes in local order parameter are more visible (figure~\ref{fig:svsp}). This suggests local order parameter as a quantity suitable to build up a phase diagram. For this purpose, we utilized the relative local order parameter, $s^*= s_f/s_i$, where  $s_i = s(0)$, $s_f = s\left(10^7\right)$.

\begin{figure}[!htbp]
\begin{minipage}[c]{0.45\textwidth}
  \centering
\includegraphics[width=\textwidth]{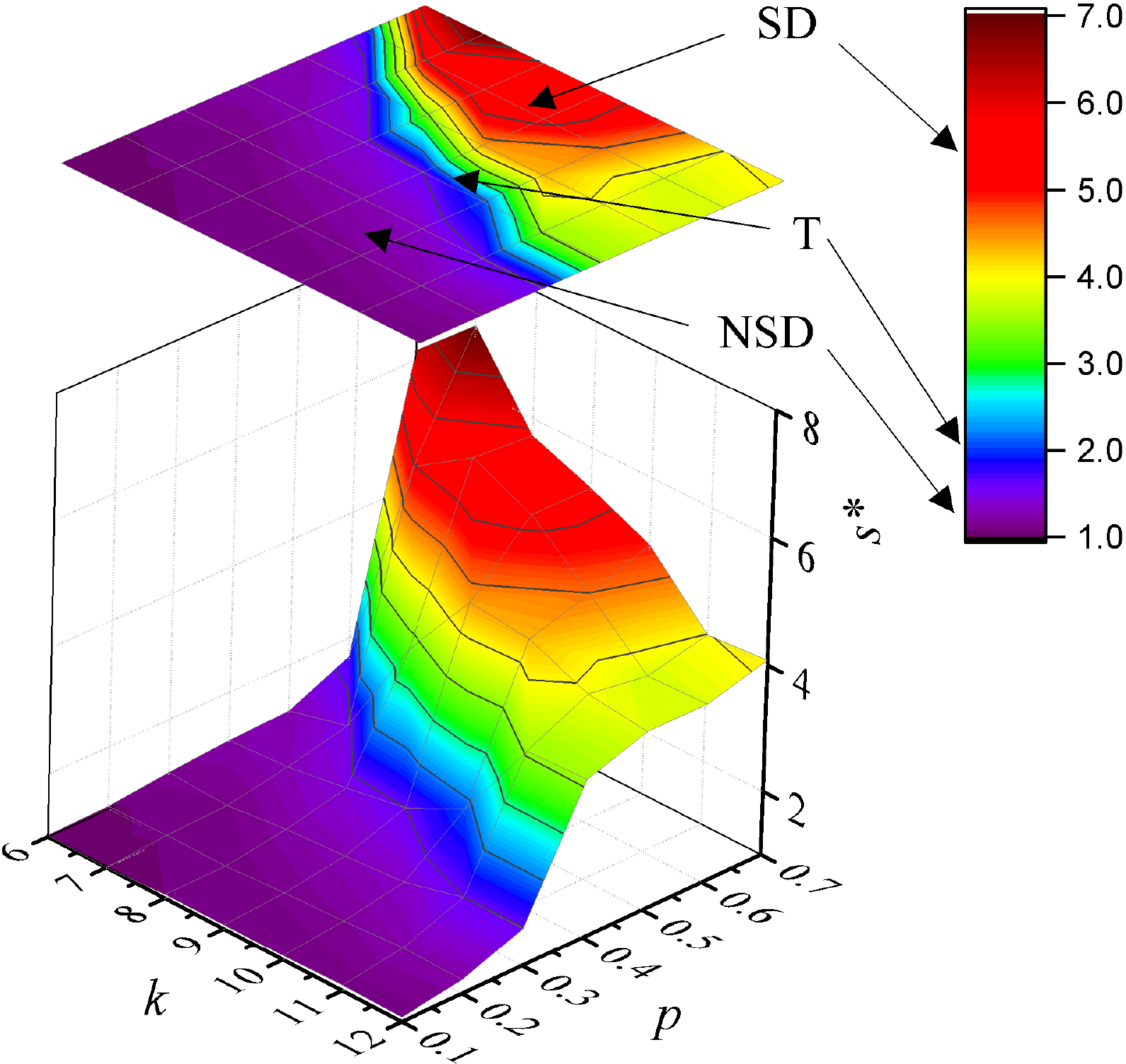}
      \end{minipage}\hfill
\begin{minipage}[c]{0.45\textwidth}
  \caption{Phase diagram in a $(k,p)$-plane; SD corresponds to the values of $k$ and $p$ when stripe domain formation occurs; NSD corresponds to the values of $k$ and $p$ when no stripe domain formation is observed; T corresponds to transient region.}\label{fig:concentrations}
  \end{minipage}
\end{figure}

Figure~\ref{fig:concentrations} presents a phase diagram for the systems under study in a ($k,p$)-plane. In the phase diagram, the region with diagonal stripe formation in the steady state at $t_{MC} \gtrsim 10^7$ is marked as SD (this region is colored from red to light blue in online version);  region corresponds to absence of stripe domains is marked as NSD (this region is colored in lilac in online version ); transient region is marked as T (this region is shown in dark blue in online version). The diagonal stripes are not observed at small packing fractions below some critical value, $p_{down}$. This critical value, $p_{down}$, continuously decreases  with increased value of $k$.

\begin{figure}[!hb]
  \centering
  \includegraphics[width=\textwidth]{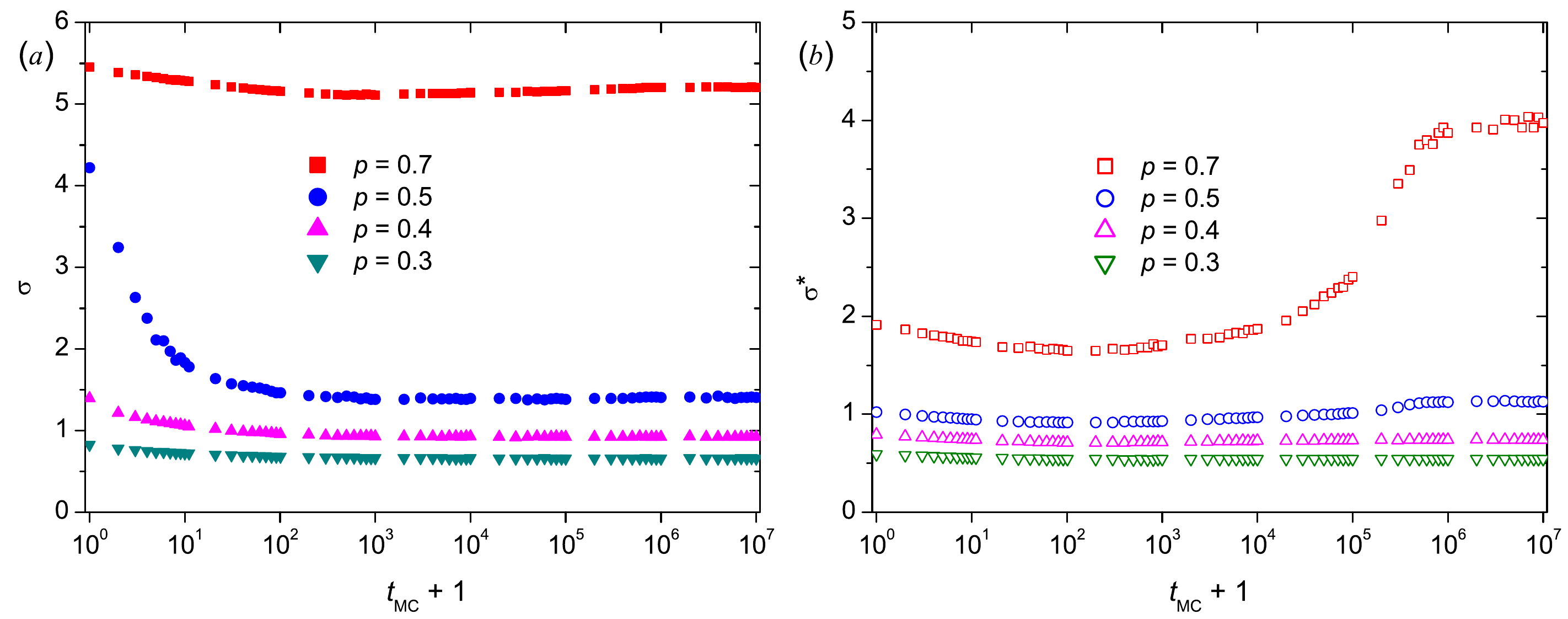}
  \caption{Examples of the temporal dynamics of electrical conductivity for $k=10$ and different values of packing fraction, $p$, $L=256$. ($a$)~Conventional model, ($b$)~Insulating ends model.}\label{fig:conductivityk10}
\end{figure}

Pattern formation affects the connectivity between domains, which can lead to a change in electrical conductivity. The conventional model is more suitable to describe initial reorganization of the system under consideration. When the packing fraction slightly exceeds the percolation threshold, the system in its initial state, i.e., a disordered system, is a conductor.  For this particular case, reorganization of the system leads to a decrease of the electrical conductivity due to decay of the percolation cluster. Figure~\ref{fig:conductivityk10} demonstrates the conductor--insulator phase transition for $k=10$ and $p=0.5$. This packing fraction is a little  greater than the percolation threshold, $p_c\approx 0.47$. In contrast, for $p=0.7$ (essentially greater than the percolation threshold) and $p = 0.4$ (somewhat below the percolation threshold), diffusion does not lead to a phase transition from conductor to insulator.

The insulating ends model can catch long-range pattern formation. This model demonstrates quite different behavior. For small values of packing fractions, the system is insulator during entirely time of simulation, whereas for the values of packing fractions close to the jamming concentration, there is the insulator--conductor phase transition. This transition occurs due to formation of dense stripe domains. Inside such the domains, lateral contacts of $k$-mers ensure electrical conductivity in both directions.

%

\section{Conclusion}\label{sec:conclusion}
Diffusional reorganization in two-dimensional systems of rigid rod-like particles has been simulated using a lattice approach. Rod-like particles were presented as linear $k$-mers of two mutually perpendicular orientations on a square lattice with periodic boundary conditions (torus). An initial homogeneous distribution of $k$-mers was produced using random sequential adsorption. The packing fraction, $p$, was varied in the range from 0.1 to the jamming packing fraction, $p_j$. By means of the Monte Carlo technique, translational diffusion of the $k$-mers was simulated as a random walk. In our model, a particle obligatorily moves when it has any possibility to change its location.  For $k \geq 6$, the system tends toward a well-organized nonequilibrium steady state in the form of diagonal stripes, when the packing fraction exceeds the critical value.

Additionally, we studied behavior of high-density systems ($p>p_j$) produced using random sequential adsorption with diffusion and additional adsorption. In such the systems, diffusivity is quite small. This results both in essential increasing of the relaxation time and in suppressing of reorganization. Even for $p \approx 1.02p_j$, stripe domain formation was not observed up to $10^8$ MC steps.

Remarkable, that pattern formation was absent both in dilute systems and in high-density systems. Stripe domain formation was observed only for systems of intermediate density. There is a curious analogy with social systems, viz., a social self-organization is possible when average number of social links per person is large enough (high population density) and individual freedom is simultaneously ensured.

\begin{figure}[htbp]
  \centering
\includegraphics[width=0.65\textwidth]{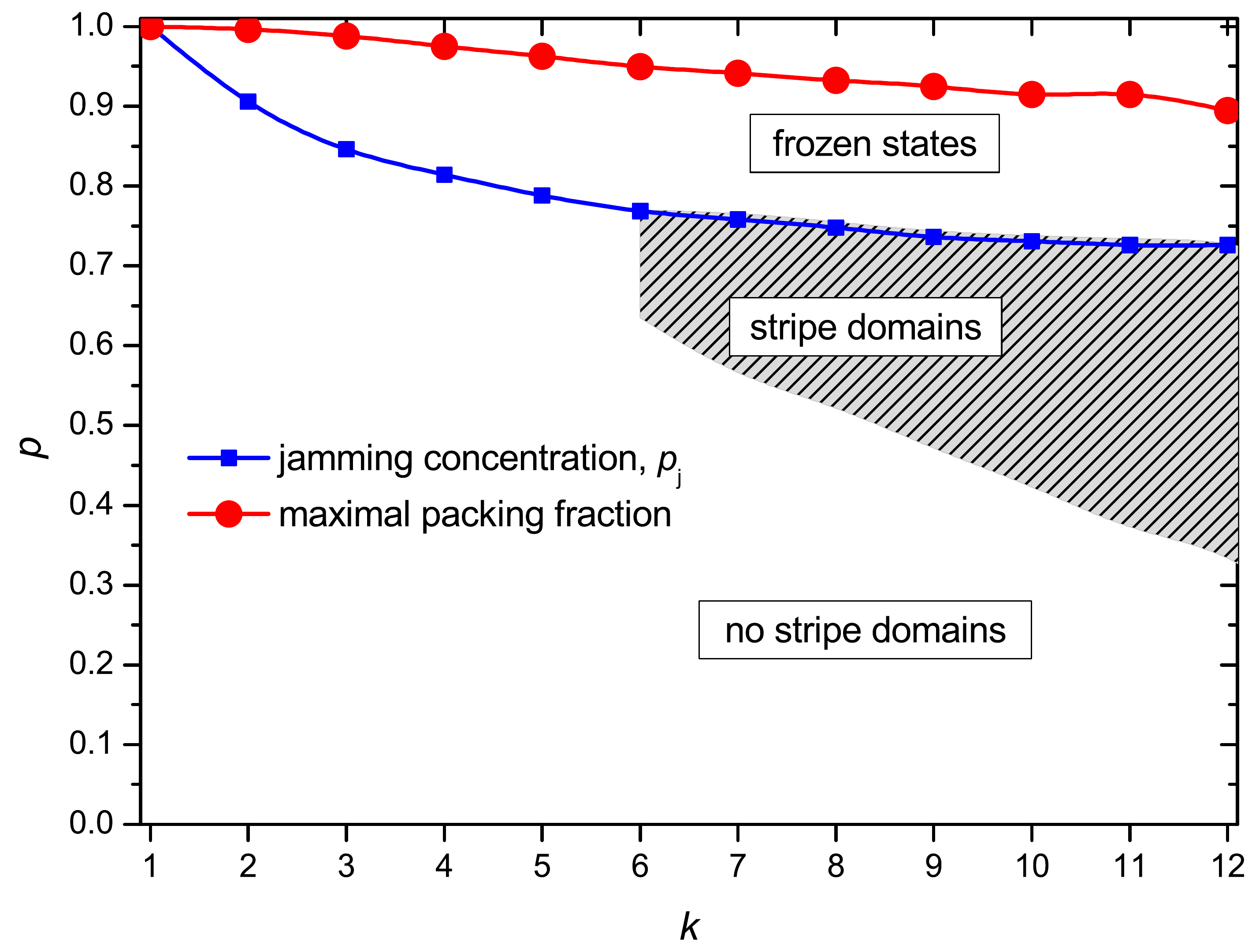}
  \caption{Phase diagram in a $(k,p)$-plane.}\label{fig:phaseplane}
\end{figure}

Figure~\ref{fig:phaseplane} presents the phase diagram obtained using both the models.


\appendix
\section*{Appendix}
\begin{figure}[htbp]
  \centering
\includegraphics[width=0.7\textwidth]{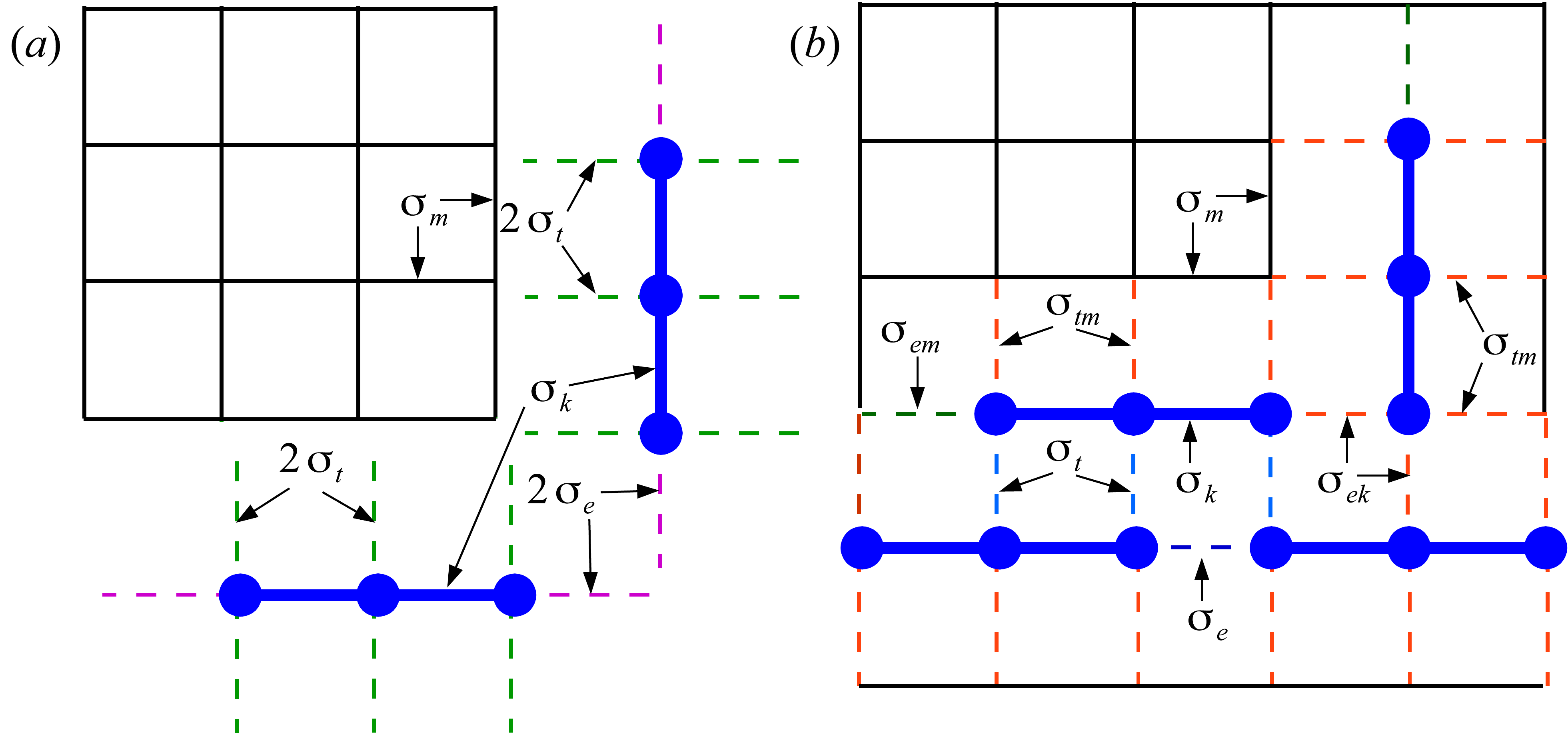}
  \caption{Transformation of the lattice with $k$-mers into RRN ($k=3$).}\label{fig:conduct}
\end{figure}

\setcounter{section}{1}
To transform system under consideration into RRN, we associate each bond between two empty sites of the lattice with the conductivity $\sigma_m$, each bond between two sites belonging to the same $k$-mer with the conductivity $\sigma_k$, each bond perpendicular to the $k$-mer with the conductivity $2\sigma_t$, and each end bond of the $k$-mer with the conductivity $2\sigma_e$ (figure~\ref{fig:conduct}($a$)). Resulting conductivities between two different sites $i$ and $j$ are calculated as
$$
\sigma_{ij} = \frac{2 \sigma_i \sigma_j}{\sigma_i + \sigma_j}.
$$
All possible combinations are presented in figure~\ref{fig:conduct}($b$).

In so-called conventional model, $\sigma_k = \sigma_t= \sigma_e = 10^6$ a.u. and $\sigma_m = 1$ a.u.
In so-called insulating ends model, $\sigma_k = \sigma_t= 10^6$ a.u. and $\sigma_m = \sigma_e = 1$ a.u.

\ack
The reported study was supported by the Ministry of Education and Science of the Russian Federation, Project No.~3.959.2017/4.6 (Y.Y.T., V.V.L., and V.V.C.), and the National Academy of Sciences of Ukraine, Project No.~43/17-H (N.I.L.).

\section*{References}
\bibliographystyle{iopart-num}
\bibliography{diffusion,mypubs}

\providecommand{\newblock}{}
\begin{thebibliography}{10}
\expandafter\ifx\csname url\endcsname\relax
  \def\url#1{{\tt #1}}\fi
\expandafter\ifx\csname urlprefix\endcsname\relax\def\urlprefix{URL }\fi
\providecommand{\eprint}[2][]{\url{#2}}

\bibitem{Narayan2006}
Narayan V, Menon N and Ramaswamy S 2006 {\em J. Stat. Mech. Theor. Exp.\/}
  {\bf 2006} P01005 ISSN 1742-5468

\bibitem{Galanis2006}
Galanis J, Harries D, Sackett D~L, Losert W and Nossal R 2006 {\em Phys. Rev.
  Lett.\/} {\bf 96} 028002 ISSN 0031-9007

\bibitem{Aranson2007}
Aranson I~S, Volfson D and Tsimring L~S 2007 {\em Phys. Rev. E\/} {\bf 75}
  051301 ISSN 1539-3755

\bibitem{Galanis2010}
Galanis J, Nossal R, Losert W and Harries D 2010 {\em Phys. Rev. Lett.\/} {\bf
  105} 168001 ISSN 0031-9007

\bibitem{Yadav2012EPJE}
Yadav V and Kudrolli A 2012 {\em Eur. Phys. J. E\/} {\bf 35} 104 ISSN 1292-895X

\bibitem{Boerzsoenyi2013}
B{\"o}rzs{\"o}nyi T and Stannarius R 2013 {\em Soft Matter\/} {\bf 9}
  7401--7418 ISSN 1744-6848

\bibitem{Muller2015PRE}
M\"uller T, de~las Heras D, Rehberg I and Huang K 2015 {\em Phys. Rev. E\/}
  {\bf 91}(6) 062207 ISSN 2470-0045

\bibitem{Gonzalez2017SM}
Gonz\'{a}lez-Pinto M, Borondo F, Mart\'{\i}nez-Rat\'{o}n Y and Velasco E 2017
  {\em Soft Matter\/} {\bf 13}(14) 2571--2582 ISSN 1744-6848

\bibitem{Aranson2006RMP}
Aranson I~S and Tsimring L~S 2006 {\em Rev. Mod. Phys.\/} {\bf 78}(2) 641--692
  ISSN 0034-6861

\bibitem{Evans1993RMP}
Evans J~W 1993 {\em Rev. Mod. Phys.\/} {\bf 65}(4) 1281--1329 ISSN 0034-68611

\bibitem{Centres2015JStatMech}
Centres P~M and Ramirez-Pastor A~J 2015 {\em J. Stat. Mech. Theor. Exp.\/} {\bf
  2015} P10011 ISSN 1742-5468

\bibitem{Kuriata2016}
Kuriata A, Polanowski P and Sikorski A 2016 {\em Macromol. Theor. Simul.\/}
  {\bf 25} 360--368 ISSN 1521-3919

\bibitem{Budinski2017PRE}
Budinski-Petkovi\'{c} L, Lon\v{c}arevi\'{c} I, Dujak D, Kara\v{c} A,
  \v{S}\'{c}epanovi\'{c} J~R, Jak\v{s}i\'{c} Z~M and Vrhovac S~B 2017 {\em
  Phys. Rev. E\/} {\bf 95}(2) 022114 ISSN 2470-0045

\bibitem{Lebovka2017PRE}
Lebovka N~I, Tarasevich Y~Y, Gigiberiya V~A and Vygornitskii N~V 2017 {\em
  Phys. Rev. E\/} {\bf 95}(5) 052130 ISSN 2470-0045

\bibitem{Tarasevich2017JSM}
Tarasevich Y~Y, Laptev V~V, Burmistrov A~S and Lebovka N~I 2017 {\em J. Stat. Mech. Theor. Exp.\/} {\bf 2017} 093203 ISSN 1742-5468

\bibitem{Fusco2001JChPh}
Fusco C, Gallo P and~Petri A and Rovere M 2001 {\em J. Chem. Phys.\/} {\bf 114}
  7563--7569 ISSN 0021-9606

\bibitem{Patra2017}
Patra S, Das D, Rajesh R and Mitra M~K 2017 {\em arXiv:1710.02283
  [cond-mat.soft]\/}

\bibitem{HK76}
Hoshen J and Kopelman R 1976 {\em Phys. Rev. B\/} {\bf 14}(8) 3438--3445 ISSN
  2469-9950

\bibitem{Frank1988PRB}
Frank D~J and Lobb C~J 1988 {\em Phys. Rev. B\/} {\bf 37}(1) 302--307 ISSN
  2469-9950

\bibitem{Tarasevich2012PRE}
Tarasevich Y~Y, Lebovka N~I and Laptev V~V 2012 {\em Phys. Rev. E\/} {\bf
  86}(6) 061116 ISSN 2470-0045

\end{thebibliography}

\end{document}